\documentclass[twoside]{article}
\usepackage[accepted]{aistats2016}

\usepackage[T1]{fontenc}
\usepackage{tgtermes}
\usepackage{amsmath}

\usepackage[scaled=0.92]{PTSans}
\usepackage{amssymb}
\newcommand{\mathbold}[1]{\ensuremath{\boldsymbol{\mathbf{#1}}}}
\newcommand{\mbf}[1]{\ensuremath{\boldsymbol{\mathbf{#1}}}}
\newcommand{\hmbtheta}{\mathbold{\widehat{\theta}}}
\newcommand{\hb}{\mathbold{\widehat{b}}}
\newcommand{\hB}{\mathbold{\widehat{B}}}
\newcommand{\hbP}{\mathbold{\widehat{P}}}

\usepackage{scalefnt,letltxmacro}
\LetLtxMacro{\oldtextsc}{\textsc}
\renewcommand{\textsc}[1]{\oldtextsc{\scalefont{1.10}#1}}

\usepackage[ttdefault=true]{AnonymousPro}

\usepackage[english]{babel}
\usepackage[parfill]{parskip}
\usepackage{afterpage}
\usepackage{framed}
\usepackage{xspace}

{\endMakeFramed}

\usepackage{lineno}

\usepackage{ragged2e}

\newcounter{parcount}

\DeclareRobustCommand{\PP}{\textcolor{Plum}{\P} }

\usepackage[usenames,dvipsnames]{xcolor}
\definecolor{shadecolor}{gray}{0.9}

\newcommand{\green}[1]{\textcolor{OliveGreen}{#1}}
\newcommand{\blue}[1]{\textcolor{MidnightBlue}{#1}}

\usepackage{graphicx}
\usepackage[labelfont=bf]{caption}
\usepackage[format=hang]{subcaption}

\usepackage{booktabs, array}

\usepackage[algoruled]{algorithm2e}
\usepackage{listings}
\usepackage{fancyvrb}
\fvset{fontsize=\small}

\usepackage{natbib}

\usepackage[colorlinks,linktoc=all]{hyperref}
\usepackage[all]{hypcap}
\hypersetup{citecolor=MidnightBlue}
\hypersetup{linkcolor=MidnightBlue}
\hypersetup{urlcolor=MidnightBlue}

\newcommand{\myeqp}[1]{\hyperref[eq:#1]{Equation~\ref*{eq:#1}}}
\newcommand{\mysec}[1]{\hyperref[sec:#1]{Section~\ref*{sec:#1}}}
\newcommand{\mysub}[1]{\hyperref[sub:#1]{Section~\ref*{sub:#1}}}
\newcommand{\mytable}[1]{\hyperref[table:#1]{Table~\ref*{table:#1}}}
\newcommand{\myfig}[1]{\hyperref[fig:#1]{Figure~\ref*{fig:#1}}}
\newcommand{\myappendix}[1]{\hyperref[appendix:#1]{Appendix~\ref*{appendix:#1}}}
\newcommand{\myalg}[1]{\hyperref[alg:#1]{Algorithm~\ref*{alg:#1}}}
\newcommand{\mytheorem}[1]{\hyperref[theorem:#1]{Theorem~\ref*{theorem:#1}}}
\newcommand{\myfootnote}[1]{\hyperref[footnote:#1]{Footnote~\ref*{footnote:#1}}}

\usepackage
[acronym,smallcaps,nowarn,section,nonumberlist]{glossaries}
\glsdisablehyper{}

\lstdefinestyle{mystyle}{
    commentstyle=\color{OliveGreen},
    numberstyle=\tiny\color{black!60},
    stringstyle=\color{BrickRed},
    basicstyle=\ttfamily\scriptsize,
    breakatwhitespace=false,
    breaklines=true,
    captionpos=b,
    keepspaces=true,
    numbers=none,
    numbersep=5pt,
    showspaces=false,
    showstringspaces=false,
    showtabs=false,
    tabsize=2
}
\newacronym{GMM}{gmm}{\emph{generalized method of moments}}
\newacronym{HMM}{hmm}{hidden Markov model}
\newacronym{MLE}{mle}{maximum likelihood estimate}
\newacronym{MAP}{map}{maximum a posteriori}
\newacronym{SVD}{svd}{singular value decomposition}
\newacronym{PSR}{psr}{predictive state representation}

\DeclareRobustCommand{\mb}[1]{\ensuremath{\boldsymbol{\mathbf{#1}}}}

\DeclareMathOperator*{\argmin}{arg\,min}

\newcommand{\mbX}{\mbf{X}}

\newcommand{\mbv}{\mathbold{v}}

\newcommand{\mbI}{\mbf{I}}

\newcommand{\mbtheta}{\mathbold{\theta}}
\newcommand{\mbTheta}{\mathbold{\Theta}}

\newcommand{\mbpi}{\mathbold{\pi}}

\newcommand{\W}{\mb{W}}
\newcommand{\bP}{\mb{P}}
\newcommand{\T}{\mb{T}}
\renewcommand{\O}{\mb{O}}
\newcommand{\B}{\mb{B}}
\renewcommand{\b}{\mb{b}}
\newcommand{\A}{\mb{A}}
\newcommand{\mbM}{\mb{M}}
\newcommand{\mbA}{\mb{A}}
\newcommand{\R}{\mb{R}}
\renewcommand{\S}{\mb{S}}
\newcommand{\U}{\mb{U}}
\newcommand{\mbg}{\mb{g}}

\newcommand{\mbzero}{\mb{0}}
\newcommand{\mbV}{\mb{V}}
\newcommand{\mbG}{\mb{G}}
\newcommand{\mbm}{\mb{m}}

\newcommand{\thetagmm}{\hmbtheta^{\text{gmm}}}

\newcommand{\thetam}{\hmbtheta^{\textsc{m}}}
\newcommand{\thetaspec}{\hmbtheta^{\text{spec}}}
\newcommand{\thetamle}{\hmbtheta^{\textsc{mle}}}
\newcommand{\thetastar}{\mbtheta^*}

\usepackage{tikz}
\usetikzlibrary{bayesnet}

\pgfdeclarelayer{edgelayer}
\pgfdeclarelayer{nodelayer}
\pgfsetlayers{edgelayer,nodelayer,main}

\definecolor{hexcolor0xbfbfbf}{rgb}{0.749,0.749,0.749}

\tikzset{>=latex}
\tikzstyle{none}   = [inner sep=0pt]

\tikzstyle{line}  = [ - ]
\tikzstyle{arrow}  = [ ->, shorten <=1pt, shorten >=1pt ]
\tikzstyle{ardash} = [ dotted, ->, shorten <=1pt, shorten >=1pt ]

\tikzstyle{empty}=[circle,opacity=0.0,text opacity=1.0,inner sep=0pt,minimum
width=0pt,minimum height=0pt]
\tikzstyle{box}=[rectangle,fill=White,draw=Black]
\tikzstyle{filled}=[circle,fill=hexcolor0xbfbfbf,draw=Black]
\tikzstyle{hollow}=[circle,fill=White,draw=Black]
\tikzstyle{param}=[rectangle,fill=Black,draw=Black,inner sep=0pt,minimum width=4pt,minimum height=4pt]

\bibpunct{(}{)}{;}{a}{,}{,}
\usepackage{pifont}
\usepackage{amsthm}
\newtheorem{theorem}{Theorem}
\newtheorem{lemma}[theorem]{Lemma}
\newtheorem{proposition}[theorem]{Proposition}

\begin{document}

\twocolumn[
\aistatstitle{Spectral M-estimation with Applications to Hidden Markov Models}
\aistatsauthor{ Dustin Tran \And Minjae Kim \And Finale Doshi-Velez }
\aistatsaddress{ Harvard University \And Harvard University \And Harvard University }
]

\begin{abstract}
Method of moment estimators exhibit appealing statistical properties,
such as asymptotic unbiasedness, for nonconvex problems.  However,
they typically require a large number of samples and are extremely
sensitive to model misspecification.  In this paper, we apply the
framework of {M-estimation} to develop both a generalized method of
moments procedure and a principled method for regularization.  Our
proposed M-estimator obtains optimal sample efficiency rates (in the
class of moment-based estimators) and the same well-known rates on
prediction accuracy as other spectral estimators.  It also makes it
straightforward to incorporate regularization into the sample moment
conditions.  We demonstrate empirically the gains in sample efficiency
from our approach on hidden Markov models.
\end{abstract}

\section{Introduction}
\label{sec:introduction}
Developing expressive latent variable models is a fundamental task in
statistics and machine learning. However, performing parameter
estimation with statistical guarantees remains challenging; in
practice, optimization techniques such as the EM
algorithm \citep{dempster77maximum} are used to find local solutions
to approximate the \gls{MLE} or \acrlong{MAP} solution.

Recently, inference techniques based on the method of moments
\citep{pearson1894contributions}, coined as \emph{spectral learning},
have gained interest because they provide consistent estimators for
many classes of models, such as \acrlongpl{HMM}
\citep{hsu2012spectral}, predictive state representations
\citep{boots2010closing}, latent tree models
\citep{parikh2011spectral}, weighted automata
\citep{balle2012spectral}, mixture models
\citep{anandkumar2014btensor}, and mixed membership stochastic
blockmodels \citep{anandkumar2014tensor}.  Spectral methods operate by
deriving low-order moment conditions on the model---such as the mean
and covariance---and matching these to moments of the observed data.
Often this moment-matching process can be solved efficiently with
linear algebra routines and can allow for parameter recovery in
settings where row-level data is unwieldy to work with (e.g. streaming
data) or unavailable (e.g. an institution may only be willing to
release summary statistics).

However, current spectral methods are extremely sensitive to
poorly-estimated moments and model misspecification.  The former
problem can be addressed, in part, by robust estimation methods of
covariances~\citep{negahban2011estimation}---though robust
estimation for higher order moments remains an open challenge.  When
the rank of the model is set too low---a form of model
misspecification---\citet{kulesza2014lowrank} demonstrate that naive
methods can lead to arbitrarily large prediction error. In
practice, there are many occasions where we may wish to learn a
low-rank approximation to a complex system.

In contrast, parameters learned from maximum likelihood and
other optimization-based estimators are robust
(assuming global optimum),
as they minimize the Kullback-Leibler divergence from the considered
model class to the true data distribution \citep{white1982maximum} and
can in certain cases achieve consistency \citep{gourieroux1984pseudo}.
With finite samples, optimization-based estimators can
achieve reasonable variances \citep{godambe1960optimum}.

Is such robustness possible for spectral methods?  Errors due to both
poor moment estimates and model misspecification can be viewed as
forms of overfitting.  Various heuristics such as early stopping are
considered in the literature \citep{mahoney2011implementing}, but they
fundamentally break assumptions for the statistical guarantees, and
are difficult to rigorously characterize; this leads to a disparity
between theory and practice.

In this paper, we analyze spectral methods from their traditional--and
more general---setting as an \emph{M-estimator}.  M-estimation has
deep roots in robust statistics (see, e.g.,
\citet{huber2009robust}). This connection emphasizes the relationship
of spectral methods to well-established alternatives such as
maximum likelihood. We use
this connection to recover the desired properties---sample efficiency
and balanced fitting. Specifically, our work makes the following
contributions:

\textbf{Provably optimal sample efficiency with respect to the
  moments}.  With the choice of weighted Frobenius norm as a metric on
the moment conditions, the M-estimation procedure corresponds to the
\gls{GMM}, whose estimator is proven to be \emph{statistically
  efficient} with respect to the information stored in the
moments. Most practically, the \gls{GMM} is sample efficient and is
thus more adaptive to scenarios where the size of the data set is
small to moderate or the data collection process results in imbalanced
samples for estimation.

\textbf{Principled regularization for sparse estimation}.  The setting
of M-estimation is naturally conducive to penalization in order to
regularize parameters, and it is commonly applied to perform
robust estimation and variable selection
\citep{owen2007robust,lambert2011robust,li2011nonconcave}. From the
Bayesian perspective, this can be interpreted as placing priors on the
parameters of interest, and where the log-likelihood is replaced by a
more general, robust, function of the data and parameters. The
proposed M-estimator automatically preserves the same bounds on the
predictive accuracy as other spectral algorithms, while also achieving
statistical efficiency.

We focus on the application of spectral M-estimation to \acrlongpl{HMM} in our
development of the theory (section ~\ref{sec:spectral}) and empirical
evaluation (section~\ref{sec:experiments}); we discuss extensions to
other latent variable models in section~\ref{sec:discussion}.

\section{Background}
\label{sec:background}
\subsection{M-estimation}
\label{sec:background:m-estimation}
We first review M-estimation \citep{huber1973robust,
van2000asymptotic}, which naturally generalizes the moment matching
used in spectral methods. Let observations $\mbX_1,\ldots,\mbX_N \in
\mathcal{X}$ be generated from a distribution with unknown parameters
$\thetastar\in\Theta$. Consider minimizing the criterion
\begin{equation*}
M_N(\mbtheta) = \sum_{n=1}^N m(\mbX_n, \mbtheta),
\end{equation*}
where $m(\cdot, \cdot):\mathcal{X}\times\Theta\to\mathbb{R}$ are
called the estimating functions
\citep{godambe1976conditional,godambe1991estimating}.
The argument
$\thetam$ which minimizes the criterion is termed the
\emph{M-estimator}.
Similarly, one may also consider \emph{penalized M-estimation} in
which one minimizes the criterion
\begin{equation}
M_N(\mbtheta) = \sum_{n=1}^N m(\mbX_n, \mbtheta) + \lambda P(\mbtheta)
,
\label{eq:penalized_m_estimation}
\end{equation}
where $m(\cdot,\cdot)$ is as before, $\lambda\in\mathbb{R}$ is fixed,
and $P(\cdot):\Theta\to\mathbb{R}$ is a specified penalty function on
the parameters.

Let $M(\mbtheta)=\mathbb{E}[m(\mbX , \mbtheta )]$. The
M-estimator $\thetam$ is consistent in that $\frac{1}{N} M_N(\mbtheta)$ uniformly
converges in probability to $M(\mbtheta)$ as $N\to\infty$, and
$\thetam$ converges to $\thetastar$ (or the closest projection, if
$\thetastar$ is not among the considered models).
In the case of
penalization, the intuition is that in the limit, the penalty term
$P(\thetastar)$ is dominated by the confidence one has from the data
(as the first summation grows with $N$).

\subsection{Generalized method of moments}
\label{sec:background:generalized}

A particular case of M-estimation is the \acrfull{GMM}, developed in the econometrics literature \citep{burguete1982unification,hansen1982large}. Given a vector-valued
function $m(\cdot,\cdot):\mathcal{X}\times\Theta\to\mathbb{R}^k$, the
\emph{moment conditions} form
\begin{equation*}
M(\thetastar) = \mathbb{E}[m(\mbX,\thetastar)] = \mbzero,
\end{equation*}
where the expectation is taken with respect to the data distribution
on $\mbX$.
In practice, we use empirical estimates of the $k$ moment
conditions using data, $\sum_{n=1}^N
m(\mbX_n,\mbtheta)$.%
\footnote{%
To simplify presentation, $m(\cdot,\cdot)$ is written as
vector-valued to connect to moment estimation. Some simple swapping of
symbols can recover the scalar-valued notation in M-estimation.%
}

In the setting where $k > |\Theta|$, the problem is overspecified and no root solution exists. One may best hope to find the set of parameters
$\thetastar$ which minimizes $\| \mathbb{E}[
m(\mbX,\mbtheta)]\|$ for some choice of
norm $\|\cdot\|$.  The \gls{GMM} estimator $\thetagmm$ is given by
minimizing a weighted criterion function,
\begin{equation}
M_N(\mbtheta)
=
\Big\|\sum_{n=1}^N m(\mbX_n,\mbtheta)\Big\|_{\W}^2
,
\label{eq:gmm}
\end{equation}
where for a positive definite matrix $\W\in\mathbb{R}^{k\times k}$,
the \emph{weighted norm} is
$
\| \mbv \|_{\W}^2 = \mbv^\top \W \mbv
$
for $\mbv\in\mathbb{R}^k$.

Under standard assumptions, the estimator $\thetagmm$ is consistent
and asymptotically normal. Moreover, if we set $\W\propto
\mathbb{E}[m(\mbX_n,\thetastar)m(\mbX_n,\thetastar)^\top]^{-1}$, then
$\thetagmm$ is statistically efficient in the class of consistent and
asymptotically normal estimators \emph{conditional on the moment
  conditions}. Therefore, if the moment conditions form a sufficient
statistic of the data (as in the \gls{MLE}), then the \gls{GMM}
estimator is optimal in that
its variance asymptotically achieves the optimal Cram\'{e}r-Rao lower
bound. More generally, the \gls{GMM} estimator achieves the Godambe
information.

One can reformulate many, if not all, examples of spectral learning
algorithms as special cases of M-estimation, and thus one can recover
the set of parameters with maximal sample efficiency using the
\gls{GMM} estimator (\myeqp{gmm}) and achieve certain
robustness properties and regularization by sufficient penalization of
the loss (\myeqp{penalized_m_estimation}).

\subsection{Hidden Markov models}
\label{sec:background:hmm}
For the remainder of this paper, we will focus on spectral estimation
and associated statistical guarantees for
\glsreset{HMM}\glspl{HMM}---applications to other
latent variable models are discussed in \mysec{discussion}.
An \gls{HMM} is defined by a 5-tuple $\{X,H,\T,\O,\mbpi \}$ where $X$ is a
set of $n$ discrete observations, $H$ is a set of $m$ discrete hidden
states, $\mbpi\in\mathbb{R}^m$ is the initial distribution over hidden
states, and the transition $\T\in\mathbb{R}^{m\times m}$ and
observation $\O\in\mathbb{R}^{n\times m}$ operators govern the
dynamics of the system:
\begin{align*}
\T_{ij} &= \Pr(h_{t+1} = i\mid h_t=j),\\
\O_{ij} &= \Pr(x_t = i\mid h_t = j).
\end{align*}
Specifically, \glspl{HMM} assume that given the hidden state $h_t$ at time
$t$, the next state $h_{t+1}$ and the current observation $x_t$ is
independent of any history before $h_t$.

We are interested in estimating the joint probabilities
$\Pr(x_{1:t})=\Pr(x_1,\ldots,x_t)$ and the conditional probabilities
$\Pr(x_{t}\mid x_{1:t-1})$. The model parameters $(\T,\O,\mbpi)$ can
also be recovered in our setup, but directly estimating the parameters can be
unstable and requires additional assumptions such as coherence
\citep{anandkumar2014tensor, mossel2005learning}.

If $\T$ and $\O$ are full rank, and $\mbpi > 0$ for all hidden states
$h\in[m]$, then \citet{hsu2012spectral} show that the following
statistics are sufficient to consistently estimate the joint
probabilities:
\begin{align}
\bP_1&\in\mathbb{R}^n& {[\bP_1]}_i & = \Pr(x_1 = i)
,
\nonumber\\
\bP_{2,1}&\in\mathbb{R}^{n\times n}& {[\bP_{2,1}]}_{ij} & = \Pr(x_2=i,x_1=j)
,
\label{eq:obreps}
\\
\bP_{3,x,1}&\in\mathbb{R}^{n\times n}& {[\bP_{3,x,1}]}_{ij} & =
\Pr(x_3=i,x_2=x,x_1=j)
,
\nonumber
\end{align}
where $\bP_{3,x,1}$ is written for all $x\in[n]$. We term these statistics
\emph{observable}, as they can be estimated directly using triplets of the
observations.

Specifically, \citet{hsu2012spectral} define the spectral model
parameters $(\b_1^\text{spec},\b_\infty^\text{spec},\B_x^\text{spec})$
as follows. Let $\U\in\mathbb{R}^{n\times m}$ be a matrix such that
$\U^\top\O$ is invertible---typically, it is the left singular vectors
corresponding to the $m$ largest singular values of $\bP_{2,1}$---and
set
\begin{align}
\label{eq:hsu}
\begin{split}
\b_1^\text{spec} &= \U^\top\bP_1,\\
\b_\infty^\text{spec} &= (\bP_{2,1}^\top\U)^\dagger\bP_1,\\
\B_x^\text{spec} &= \U^\top\bP_{3,x,1}(\U^\top\bP_{2,1})^\dagger\quad \forall x\in[n]
.
\end{split}
\end{align}
where $\mb{A}^\dagger$ denotes the pseudoinverse of $\mb{A}$.
Then the joint probability satisfies
\begin{equation}
\label{eq:prob:joint}
\Pr(x_{1:t})
=
\b_\infty^{T} \B_{x_t}\cdots \B_{x_1} \b_1
.
\end{equation}
Intuitively, one can think of $\b_1$ as the initial state vector in a projected
observable representation space; the matrix $\B_x$ is an observable transition operator which
propagates changes in this space; the vector $\b_\infty$
simply acts as a normalizer.
From \myeqp{prob:joint},
\citet{hsu2012spectral} demonstrated that
the estimator
$\thetaspec=(\hb_1^\text{spec},\hb_\infty^\text{spec},\hB_x^\text{spec})$,
which is constructed from the empirical statistics
$\hbP_1,\hbP_{2,1},\hbP_{3,x,1}$, is asymptotically unbiased as the empirical
statistics become exact in the limit. Moreover, the number of observations
required to achieve a fixed level of accuracy is only polynomial in
the length of the sequence, $t$.

\section{Spectral M-estimation}
\label{sec:spectral}
Following the results of spectral methods
\citep{hsu2012spectral,boots2010closing,balle2012spectral,cohen2012spectral,arora2012learning},
it is natural to consider the underlying framework for its methodology, and how
it connects to techniques for maximum likelihood estimation.
To address this, we start by deriving the usual spectral estimator \eqref{eq:hsu} from
the M-estimation setting.
\subsection{Spectral M-estimator}
\label{sec:spectral:derivation}
\newcommand{\theoremmoments}{
}
Denote the parameter triplet $\mbtheta=(\b_1, \B,
\b_\infty)$ and define the moment conditions
\begin{align}
\label{eq:moments}
\begin{split}
m_1(\mbtheta)= \b_1 &- \bP_1
,\\
m_\infty(\mbtheta) = \bP_{2,1}^\top\b_\infty &- \bP_1
,\\
m_x(\mbtheta) = \bP_{3,x,1} &- \B_x\bP_{2,1} \quad \forall x\in[n]
.
\end{split}
\end{align}
Let $\thetastar$ denote the root solution
$m_1(\thetastar)=m_\infty(\thetastar)=m_x(\thetastar)=0$.
The vector
$\b_1$ is trivially given by
$\bP_1$, and the
solution of $\b_\infty$ to $m_\infty(\cdot)$ is simply the vector of
ones,
$\mathbf{1}_n$. Thus it suffices to estimate the tensor $\B$.

The standard approach in spectral methods (e.g.,
\citet{hsu2012spectral,boots2010closing}) is to first observe that parameter
triplets satisfying the joint probability in \myeqp{prob:joint} are
equivalent up to a similarity transform: given the triplet $(\b_1, \{\B_x\},
\b_\infty)$ and an invertible matrix $\S\in\mathbb{R}^{n\times n}$, the
transformed triplet $(\b'_1= \S \b_1, \{\B'_x=\S\B_x\S^{-1}\}, \b'_\infty=
\S^{-T} \b_\infty)$ provide the same quantities.
Thus, what we are really interested in is not a unique
set of parameters but an equivalence class---governed by the joint probability
\eqref{eq:prob:joint}---and which denote identical parameters
up to a similarity transform. The moment conditions \eqref{eq:moments} are
constructed such that the solution $\thetastar$ defines a unique element in this
equivalence class (and thus by M-estimation theory, the estimator is
identifiable \citep{van2000asymptotic}).

We now formalize the connection to the usual spectral estimator as follows.
Let $\mbX=\{\mbX_n=(x_{n1},x_{n2},x_{n3})\}$ denote the data set of $N$
triplets by which the observable representations $\bP_1$,
$\bP_{2,1}$ and $\bP_{3,x,1}$ are estimated. Define
\begin{equation}
M_{N}(\B) =
\sum_{x,i,j\in[n]^3}
([ \widehat \bP_{3,x,1} ]_{ij} - [\B_x]_{i\cdot} [\widehat \bP_{2,1}]_{\cdot
j})^2
.
\label{eq:m:original}
\end{equation}
\newcommand{\propequivalence}{
Let $\thetaspec$ denote the estimator using empirical statistics in
\myeqp{hsu}.
Let $\thetam$ denote the M-estimator given by
\begin{align*}
\widehat \b_1^\text{M} &= \widehat \bP_1
,\\
\widehat \b_\infty^\text{M} &= \mathbf{1}_n
,\\
\widehat \B^\text{M} &= \argmin_{\B\in\mathbb{R}^{n\times n\times n}} M_N(\B)
.
\end{align*}
Then $\thetam$ is in the same equivalence class
as $\thetaspec$, so they provide the same probability estimates.
}
\begin{proposition}[Equivalence]
\label{prop:equivalence}
\propequivalence
\end{proposition}
Proposition \ref{prop:equivalence} allows us to leverage both M-estimation theory and the
usual finite sample bounds on accuracy given by
\citet{hsu2012spectral}.  Specifically, the sample complexity of
$\thetam$ depends polynomially on the singular values
$1/\sigma_m(\bP_{2,1})$ and $1/\sigma_m(\O)$, where $\sigma_m(\cdot)$
denotes the $m^{th}$ largest singular value of its matrix argument.

\subsection{Regularized Spectral M-estimator: Low Rank Setting}
\label{sec:spectral:generalized}

Suppose there is a low rank constraint on the parameters, where
$\operatorname{rank}(\B_x)\le k$ for some $k<m$ and for all matrices
$\B_x$.  We may impose this constraint for computational tractability,
to avoid the $\mathcal{O}(n^3)$ complexity of solving \acrlong{SVD}
associated with the dynamical system.  It may also occur naturally:
the maximal rank of $\B_x$ is
$\operatorname{rank}(\O)=\operatorname{rank}(\T)\le m$, and often the
transition operators are low rank.
Estimation with this constraint is known as \emph{low rank spectral
  learning}, \citet{kulesza2014lowrank} show that simply truncating
$\B_x$ to a desired rank can lead to poor prediction.  Following the
M-estimation setting, we now derive a more robust estimator.

To optimize over an unconstrained Euclidean space, we first cast the
low rank estimation problem in terms of matrix factorization. Let
$\B_x=\R_x\S_x^\top$, where $\R_x,\S_x\in\mathbb{R}^{n\times k}$, and
let $\R$ and $\S$ be tensors formed by the collections of matrices
$\{\R_x\}$ and $\{\S_x\}$ respectively.

This leads to the criterion function
\begin{equation}
\label{eq:low_rank}
M_{N}(\B) =
\sum_{x,i,j\in[n]^3}
([ \widehat \bP_{3,x,1} ]_{ij} - [\R_x]_{i\cdot}\S_x^\top [\widehat \bP_{2,1}]_{\cdot
j})^2
,
\end{equation}
where we use the notation $\A_{i\cdot}$ (and respectively, $\A_{\cdot j}$) to
represent the $i^{th}$ row (and $j^{th}$ column) of a matrix.

\subsection{Regularized Spectral M-estimator: Additional Penalization}
\label{sec:spectral:penalized}
Given the M-estimation following \myeqp{low_rank}, we can generalize the procedure
further by augmenting the criterion function with a penalty term,
\begin{equation*}
M_N(\R,\S) + \lambda P_\alpha(\R,\S)
,
\end{equation*}
where $P_\alpha(\R,\S)$ is a specified penalty function with regularization
parameter $\lambda$.
However, in general, if $M_N(\R,\S)$ converges in probability to $M(\R,\S)$ as
in the current setting, we must specify a suitable decaying schedule on the
penalty function,
\begin{equation*}
M_N(\R,\S) + \lambda N^{-p} P_\alpha(\R,\S)
\end{equation*}
for fixed $p>0$ (unlike traditional penalized M-estimation, the number
of summations remains fixed as $N\to\infty$). Ideally, the penalty
function should decay at the slowest possible rate, without affecting
the convergence rate of the previous M-estimator
\eqref{eq:low_rank}. We choose $p$ as follows.

\newcommand{\propositiondecay}{
Let $\thetam$ denote the M-estimator obtained by minimizing the criterion
function
\begin{equation*}
M_N(\R,\S) + \lambda N^{-p} P_\alpha(\R,\S),
\end{equation*}
where $p>0$.
Then the largest value of $p$ such that the convergence rate of $\thetam$
does not change is $p=1/2$.
}
\begin{proposition}
\label{proposition:decay}
\propositiondecay
\end{proposition}
Trivially this is the case based on the asymptotic rate of the estimator.
In practice, we consider losses of the form
\begin{equation}
\label{eq:loss}
\mathcal{L}(\R,\S) = M_N(\R,\S) + \lambda N^{-1/2} \|\R\|_1
.
\end{equation}
Penalizing only the first factor of $\B$ acts as a proxy for
penalizing the observation operator $\O$; that is, by construction one
can show that $\B_x=\O\mbA_x \O^\dagger$, where
$\mbA_x=\T\operatorname{diag}(\O_{x,1},\ldots,\O_{x,m})$.  We will
denote this final criterion function as $\mathcal{L}$ and its
M-estimator as $\thetam$, which also collects the two parameters
$\hb^\textsc{m}_1=\hbP_1$ and $\hb^\textsc{m}_\infty=\mb{1}$.

\subsection{Sample Efficiency through Generalized Method of Moments}
With the low rank and penalization extensions in place, we extend
the estimation procedure once more: we define the criterion
function $M_N(\R,\S)$ of \myeqp{loss} in order to obtain optimal
sample efficiency.

Let $\mbm$ be a vector of length $n^3$, which flattens
the moment conditions $m_x(\mbtheta)$ over $x\in[n]$ and
each matrix element $i,j$. More specifically, an index
$(x,i,j)\in[n]^3$ into $\mbm$ is
\begin{equation*}
\mbm_{xij} =
[ \widehat \bP_{3,x,1} ]_{ij} - [\R_x]_{i\cdot}\S_x^\top [\widehat \bP_{2,1}]_{\cdot j}
.
\end{equation*}
As before, there are $n^3$
moment conditions but now $2n^2k$
parameters due to the low rank structure---corresponding to each
element in the
$n\times n\times k$ tensors $\R,\S$.
The \gls{GMM} estimator is the minimizer of the criterion function
\begin{equation}
\label{eq:m:gmm}
M_N(\R,\S)
=
\sum_{i,j\in[[n]^3]^2}
\W_{ij} \mbm_i \mbm_j
,
\end{equation}
where $\W$ is a weighting matrix that trades off between errors in the
various \gls{GMM} moment condtions.  If $\W$ is the identity $\mbI$,
then each term is $\mbm_i\mbm_j$ for all $i,j\in[n]^3$;
this recovers the original spectral M-estimation criterion function
considered in \myeqp{low_rank}.

To achieve maximum sample efficiency, \gls{GMM} theory
\citep{hansen1982large} states that the optimal weighting $\W$ is
proportional to the precision matrix,
\begin{equation}
\label{eq:weighting}
\W\propto
\mathbb{E}[m(\mbX_n,\{\R^*,\S^*\})m(\mbX_n,\{\R^*,\S^*\})^\top]^{-1}
.
\end{equation}
The optimal
$\W$ minimizes the variance of the estimator by calibrating
it to the inexactness of the estimated statistics, $\hbP_1,\hbP_{2,1},\hbP_{3,x,1}$.
If the moment conditions form the gradient of the log-likelihood function, $m(\mbX, \{\R,\S\}) = \nabla
\ell(\{\R,\S\}; \mbX)$,
the optimal weighting matrix $\W$ becomes
the inverse Fisher information evaluated at the true parameters. This
recovers a maximum likelihood estimator with minimal asymptotic
variance. Analogous to the \gls{MLE} setting, the choice of
the moment conditions $m$ and weighting matrix $\W$ may also be
interpreted as minimizing a distance to the true data generating
distribution, where the distance between probability distributions is
defined by symmetrized KL divergence
\citep{amari1997information,amari1997estimating}.

To gain intuition, note that a first-order diagonal approximation to
\myeqp{weighting} is given by the inverse diagonal entries of the
expected outer product.  These entries $\W_{ii}$ weight according to the
magnitude of error in the sample moments $\mbm_{i}$. Large magnitudes
for $\mbm_{i}$ lead $1/\mbm_{i}^2$ to be small; this forces the
M-estimator to place less weight on high error moments.  With
cross-correlation, $\W$ places more weight on other estimates paired
with high error moments. For example, a small error moment $\mbm_j$
leads to a larger weight $1/(\mbm_i \mbm_j)^2$. These weights enable
more intelligent parameter estimation.

\section{Algorithm}
\label{spectral:algorithm}
The criterion function $\mathcal{L}$ of \myeqp{loss} is a quadratic form plus
a convex penalty. Moreover, it is strongly convex for $\R$ given $\S$ and $\S$
given $\R$. Hence we proceed with estimation by the procedure of alternating
minimization, i.e., apply convex solvers which alternate between estimating
each set of parameters.

More specifically, we apply an iterative procedure where we 1. alternate
minimizing the loss over $\R$ and $\S$ conditioned on an estimate of $\W$; 2.
set $\W$ conditioned on estimates of $\R,\S$; 3. repeat the procedure until
convergence. An overview of the procedure is
described in \myalg{hmm:m}, and we derive gradients in the
following proposition.

\newcommand{\propgradient}{
The gradients are
\begin{align}
\nabla_{\R}\mathcal{L} &=
\mathcal{J}_{\R}^\top\W
m(\mbX, \{\R,\S\})
+ \nabla_{\R} P_{\alpha}(\R,\S)
\label{eq:gradr}
\\
\nabla_{\S}\mathcal{L} &=
\mathcal{J}_{\S}^\top\W
m(\mbX, \{\R,\S\})
+ \nabla_{\R} P_{\alpha}(\R,\S)
\label{eq:grads}
\end{align}
where the matrices $\mathcal{J}_{\R}\in\mathbb{R}^{n^3\times n^2k}$ and
$\mathcal{J}_{\S}\in\mathbb{R}^{n^3\times n^2k}$ are given by
\begin{equation}
[\mathcal{J}_{\R}]_{xij, uvw}
=
\begin{cases}
-[\S_x^\top]_{w\cdot}[\bP_{2,1}]_{\cdot j}, &\text{ if }x=u,~i=v\\
0, &\text{ otherwise}
\end{cases}
\end{equation}
and
\begin{equation}
[\mathcal{J}_{\S}]_{xij, uvw}
=
\begin{cases}
-[\R_x]_{iw}[\bP_{2,1}]_{vj}, &\text{ if }x=u\\
0, &\text{ otherwise}
\end{cases}
\end{equation}
}
\begin{proposition}
\label{prop:gradient}
\propgradient
\end{proposition}

\begin{algorithm}[t]
  \caption{Spectral M-estimation for \glspl{HMM}}
  \SetAlgoLined
  \DontPrintSemicolon
  \BlankLine
  \KwIn{$N$ observation triplets $\mbX=\{\mbX_n:(x_1,x_2,x_3)\}$.}
  Construct empirical statistics $\widehat \bP_1, \widehat \bP_{2,1}$, $\widehat
  \bP_{3,x,1}~\forall x\in[n]$.\;
  Initialize $\widehat \W=I$.\;
  Set iteration counter $s = 1$.\;
  \BlankLine
  \While{not converged}{
    \If{$s\ge 2$}{
      $\widehat \W = \left(\sum_{n=1}^N m(\mbX_n, \{\widehat \R,\widehat
      \S\})m(\mbX_n, \{\widehat \R,\widehat \S\})^\top\right)^{-1}$
    }
    \BlankLine
    $\widehat \R, \widehat \S = \operatornamewithlimits{arg\, min}_{\R,
    \S}\mathcal{L}(\R,\S)$ (\myalg{hmm:gradient}).
    \BlankLine
    Increment $s$.\;
  }
  $\hb^\textsc{m}_1 = \hbP_1$.\;
  $\hB^\textsc{m} =\{ \widehat \R_x \widehat
  \S_x^\top\}$.\;
  $\hb^\textsc{m}_\infty = \mathbf{1}_n$.\;
  Return $\thetam=(\hb^\textsc{m}_1, \hB^\textsc{m}, \hb^\textsc{m}_\infty)$.\;
  \label{alg:hmm:m}
\end{algorithm}

\begin{algorithm}[t]
  \caption{Alternating minimization, given weights $\W$}
  \SetAlgoLined
  \DontPrintSemicolon
  \BlankLine
  \KwIn{initial values $\widehat \R,\widehat \S$.}
  \BlankLine
  \While{not converged}{
    $\widehat\R = \argmin_{\R} \mathcal{L}(\R,\S)$ (\myeqp{gradr})
    \BlankLine
    $\widehat\S = \argmin_{\S} \mathcal{L}(\R,\S)$ (\myeqp{grads})
  }
  Return $\widehat \R, \widehat \S$.\;
  \label{alg:hmm:gradient}
\end{algorithm}

Note that we initialize $\widehat \W=\mbI$, so that one loop of
\myalg{hmm:m} corresponds to
the original spectral estimator of \myeqp{m:original}. The global optima upon future
iterations are refined based on the weighting matrix, and are in fact
\emph{guaranteed to perform at least as well} as the minimizer of the original optima. Note
also that only one iteration of the loop is necessary for optimal sample
efficiency asymptotically, as $\widehat \W$ converges in probability to
$\mathbb{E}[m(\mbX_n,\{\R,\S\})m(\mbX_n,\{\R,\S\})^\top]^{-1}$. However, for
finite data we see in experiments that better performance occurs when running
the algorithm until convergence.

The matrix factorization view considered here, as well as the introduction of
the weighting matrix $\W$, makes the problem highly nonconvex.  However, much recent theory has gone into explaining why simple
optimization procedures following alternating minimization typically perform
well in practice
\citep{jain2013low,loh2014regularized,hardt2014understanding,chen2015fast,bhojanapalli2015dropping,garber2015fast,loh2015statistical}.
We also find that
in practice the richer information gain from the generalized
M-estimation procedure leads to improved
estimates. It is an open problem to understand these
improvements theoretically. Note that initialization using the original spectral estimator
guarantees a global solution to the first iteration without
penalization; we can apply it to initialize future iterations of the weighting as well as for
nonconvex optimization with a penalty.

For computational efficiency, one can take immediate advantage of the block
diagonal structure of
the weighting matrix: this comes as a result of the independent sets of
parameters in the loss function of \myeqp{loss}. That is, the
parameter matrices $\B_{x'}$ only appear in the $m_{xij}\in[n]^3$ moments
when $x=x'$. Thus it can be embarassingly parallelized into $n$ separate
optimizations. We apply individual optimizations on $n$
procedures, each of which have $n^2$ moment conditions and recover a particular
$\B_x$. The computational complexity of the algorithm is
$\mathcal{O}(n^2)$ per iteration, with a storage complexity of
$\mathcal{O}(n^4)$.

\vspace{-1.5ex}
\section{Experiments}
\label{sec:experiments}
\vspace{-1.5ex}
We demonstrate the sample efficiency gained by the weighting scheme in
the M-estimator and the advantage of sparse estimation due to $L_1$
penalization. We use toy configurations to highlight the M-estimator's
robustness to model or rank mismatch, imbalanced observations, low
sample size, and overfitting; finally we show results on real data.

\begin{figure}[!b]
\begin{minipage}{\textwidth}
\begin{minipage}[!t]{.3\textwidth}
\centering
\begin{tabular}{llll}
\toprule
Length & $\widehat \B_0^\text{spec}$ & $\widehat \B_0^\textsc{m}$\\
\midrule
10 & \textbf{1} & \textbf{1}\\
15 & 0.8889 & \textbf{0.8607}\\
25 & 0.0198 & $\mathbf{3.1521\cdot 10^{-6}}$\\
50 & 0.0008 & $\mathbf{9.9664\cdot 10^{-9}}$\\
\bottomrule
\end{tabular}
\label{table:sequence}
\end{minipage}
\begin{minipage}[!t]{0.6\textwidth}
\includegraphics[height=1.14in]{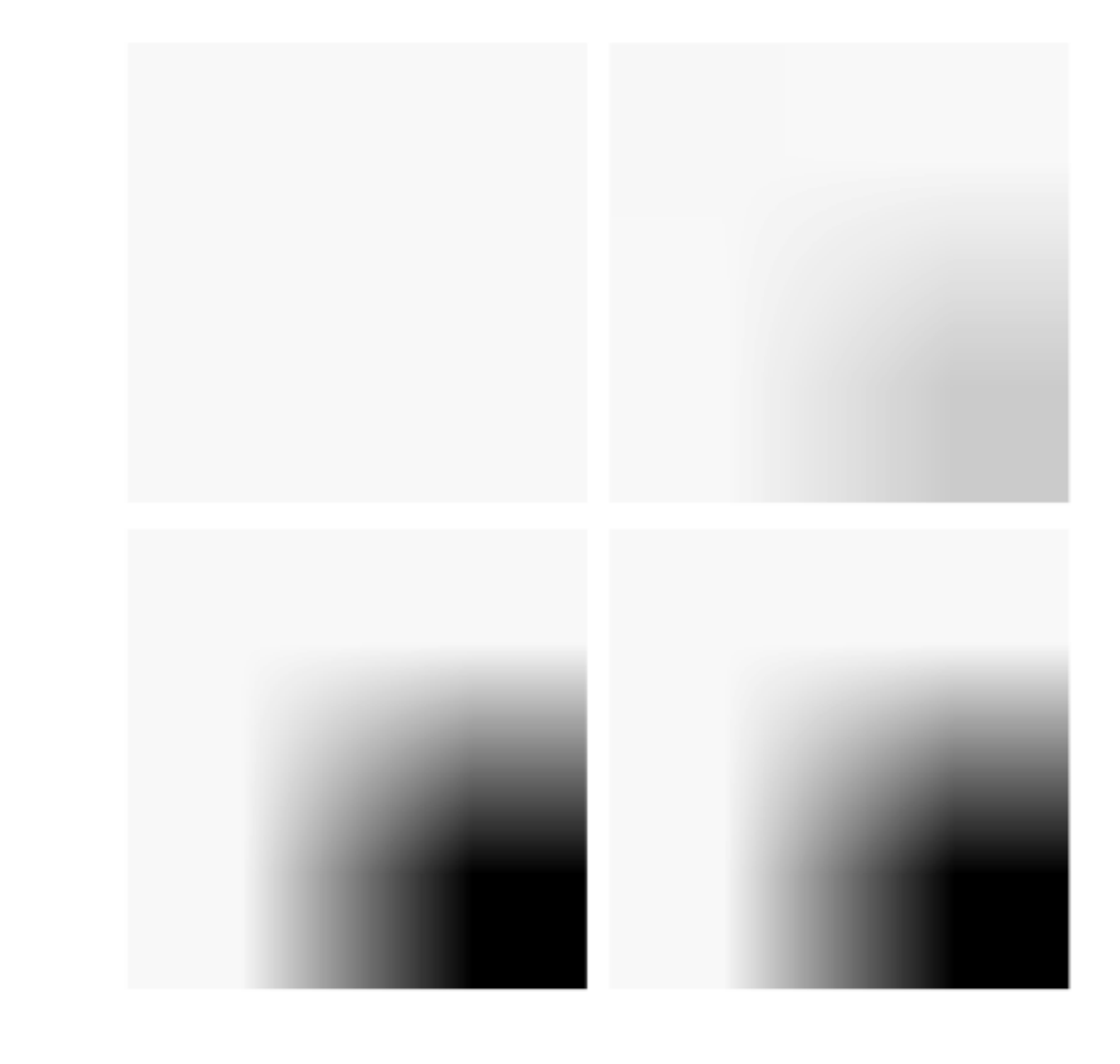}
\end{minipage}
\end{minipage}
\caption{Left: Decay of the transition operator $\B_0$ as the
length of sequence increases (lower is better);
Right: Weighting matrix of $\B_0$ for each length is displayed from top
left-right, bottom left-right.}
\label{fig:sequence}
\end{figure}

For the M-estimator, we initialize using the original spectral
estimate and also try several random initializations; we
then take the estimates with minimal training loss.  As the weighting
matrix can become numerically singular, we add $10^{-8}$ to the
diagonal.  Comparisons are always done on test set evaluations. Note
also that evaluations of the loss cannot be compared among algorithms,
as the estimators minimize inherently different functions.

\subsection{Deterministic sequence}

Consider a rank-11 system of two binary states: 0 and 1. The
observation sequence deterministically follows the pattern
"00000000001\ldots 1", where 0 is always observed for the first 10
steps and 1 is observed for all remaining steps.  Suppose that we aim
to estimate this with a model of rank 1.
Figure \ref{fig:sequence} displays the original estimator $\thetaspec$
and the M-estimator $\thetam$. As the length of the sequence
increases, we expect $\B_0$, the observable transition operator for
the first state, to decay to 0.  Our M-estimator achieves this at a
much faster rate than $\thetaspec$. It places more weight on the first
state, and this weight increases with the length of the sequence.

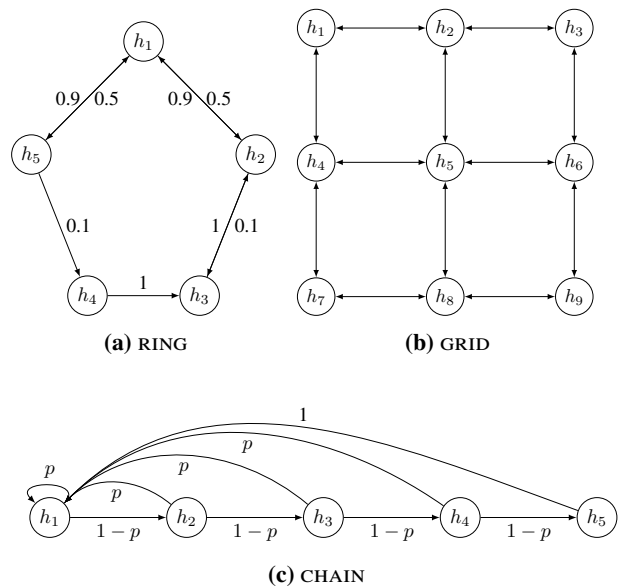
\begin{figure}[tb]
\begin{subfigure}[t]{0.45\columnwidth}
  \centering
  \resizebox {\columnwidth} {!} {
    \begin{tikzpicture}[x=1.7cm,y=1.8cm]

  \node[latent]  (h1) {$h_1$} ;
  \node[latent, below=of h1, xshift=2.0cm, yshift=0.5cm] (h2) {$h_2$} ;
  \node[latent, below=of h2, xshift=-1.0cm] (h3) {$h_3$} ;
  \node[latent, below=of h1, xshift=-2.0cm, yshift=0.5cm] (h5) {$h_5$} ;
  \node[latent, below=of h5, xshift=1.0cm] (h4) {$h_4$} ;

  \draw[->] (h1) to node[midway,above,right] {0.5} (h2);
  \draw[<-] (h1) to node[midway,below,left] {0.9} (h2);
  \draw[->] (h2) to node[midway,below,right] {0.1} (h3);
  \draw[<-] (h2) to node[midway,above,left] {1} (h3);
  \draw[<-] (h3) to node[midway,above] {1} (h4);
  \draw[->] (h1) to node[midway,below,right] {0.5} (h5);
  \draw[<-] (h1) to node[midway,above,left] {0.9} (h5);
  \draw[<-] (h4) to node[midway,above,right] {0.1} (h5);

\end{tikzpicture}
  }
  \caption{\textsc{ring}}
  \label{sub:ring}
\end{subfigure}
\begin{subfigure}[t]{0.5\columnwidth}
  \centering
  \resizebox {\columnwidth} {!} {
    \begin{tikzpicture}[x=1.7cm,y=1.8cm]

  \node[latent]  (h1) {$h_1$} ;
  \node[latent, right=of h1] (h2) {$h_2$} ;
  \node[latent, right=of h2] (h3) {$h_3$} ;
  \node[latent, below=of h1] (h4) {$h_4$} ;
  \node[latent, right=of h4] (h5) {$h_5$} ;
  \node[latent, right=of h5] (h6) {$h_6$} ;
  \node[latent, below=of h4] (h7) {$h_7$} ;
  \node[latent, right=of h7] (h8) {$h_8$} ;
  \node[latent, right=of h8] (h9) {$h_9$} ;

  \draw[<->] (h1) to node[midway,below] {} (h2);
  \draw[<->] (h2) to node[midway,below] {} (h3);
  \draw[<->] (h4) to node[midway,below] {} (h5);
  \draw[<->] (h5) to node[midway,below] {} (h6);
  \draw[<->] (h7) to node[midway,below] {} (h8);
  \draw[<->] (h8) to node[midway,below] {} (h9);
  \draw[<->] (h1) to node[midway,left] {} (h4);
  \draw[<->] (h2) to node[midway,below] {} (h5);
  \draw[<->] (h3) to node[midway,below] {} (h6);
  \draw[<->] (h4) to node[midway,left] {} (h7);
  \draw[<->] (h5) to node[midway,left] {} (h8);
  \draw[<->] (h6) to node[midway,left] {} (h9);

\end{tikzpicture}
  }
  \caption{\textsc{grid}}
  \label{sub:grid}
\end{subfigure}
\\
\begin{subfigure}[t]{\columnwidth}
  \centering
  \resizebox {\columnwidth} {!} {
    \begin{tikzpicture}[x=1.7cm,y=1.8cm]

  \node[latent]  (h1) {$h_1$} ;
  \node[latent, right=of h1] (h2) {$h_2$} ;
  \node[latent, right=of h2] (h3) {$h_3$} ;
  \node[latent, right=of h3] (h4) {$h_4$} ;
  \node[latent, right=of h4] (h5) {$h_5$} ;

  \draw[->] (h1) to node[midway,below] {$1-p$} (h2);
  \draw[->] (h2) to node[midway,below] {$1-p$} (h3);
  \draw[->] (h3) to node[midway,below] {$1-p$} (h4);
  \draw[->] (h4) to node[midway,below] {$1-p$} (h5);
  \draw[->] (h1) to[out=45,in=135,looseness=3] node[midway,above] {$p$} (h1);
  \draw[->] (h2) to[out=140,in=45] node[midway,below] {$p$} (h1);
  \draw[->] (h3) to[out=140,in=45] node[midway,below] {$p$} (h1);
  \draw[->] (h4) to[out=145,in=45] node[midway,below] {$p$} (h1);
  \draw[->] (h5) to[out=160,in=45] node[midway,above] {1} (h1);

\end{tikzpicture}
  }
  \caption{\textsc{chain}}
  \label{sub:chain}
\end{subfigure}
\vspace{-1ex}
\caption{%
\gls{HMM} configurations.
\textbf{(a)} \textsc{ring}: The outer loop indicate clockwise
transition probabilities, the inner loop indicate
counter-clockwise.
\textbf{(b)} \textsc{grid}: Each state has equal probability of visiting any neighbor.
\textbf{(c)} \textsc{chain}: States transition with a probability $p$ of
resetting to the first state.%
}
\label{fig:toy}
\end{figure}

\subsection{Ring configuration}

\begin{table}[h]
\centering
\begin{tabular}{llll}
\toprule
Model rank & $\thetaspec$ & $\thetam$ & ($\thetam$, $\lambda=0.01$) \\
\midrule
4 & 1.50 & 1.25 & \textbf{1.03}\\
3 & 1.15 & 1.03 & \textbf{0.81}\\
2 & 0.68 & 0.65 & \textbf{0.60}\\
\bottomrule
\end{tabular}
\caption{Relative norm difference between estimated and true joint
probability, averaged over 100 test examples.}
\label{table:ring}
\end{table}

In \mysub{ring}, the hidden states form a ring: $h_1$ has uniform
probability of proceeding clockwise to $h_2$ or counter-clockwise to
$h_5$; $h_2$ and $h_5$ return back to $h_1$ with probability 0.9 and
visit $h_3$ or $h_4$ (respectively) with probability 0.1.
This leads to imbalanced samples where $h_1,h_2,h_5$ are visited most,
and one rarely sees $h_3$ and $h_4$.  States are correctly observed
with 0.6 probability, otherwise we observe any other state
uniformly. We train on 100 examples.

\mytable{ring} shows that under difficult settings---with imbalanced states,
not enough training examples, and ill-posed rank problems---spectral estimators
fit poorly due to the information loss from higher order moments. However,
the weighting scheme of $\thetam$ allows the estimator to compensate for some
of these problems, and thus it performs better than $\thetaspec$. Moreover,
when used with a $L_1$ penalty of $\lambda=0.01$, the estimator dominates
other algorithms; the value of $\lambda$ was also chosen generally and not
optimized over.

\subsection{Grid configuration}

\begin{table}[!h]
\centering
\begin{tabular}{llll}
\toprule
Grid size & $\thetaspec$ & $\thetam$ & ($\thetam$,
$\lambda=1\textsc{e}\text{-}3$) \\
\midrule
$2\times 2$ & 0.014 & 0.014 & \textbf{0.014} \\
$3\times 3$ & 0.225 & 0.225 & \textbf{0.212} \\
$5\times 5$ & 0.475 & 0.475 & \textbf{0.458}\\
\bottomrule
\end{tabular}
\caption{Relative norm difference between estimated and true joint
probability, averaged over 100 test examples.}
\label{table:grid}
\end{table}

In the grid configuration (\mysub{grid}), each hidden state has an equal
probability of transitioning to any one of its neighbors; the observation
matrix $\O$ indicates the correct state with 0.9 probability, and any other
state otherwise. We use 100,000 training examples and vary the grid
size.

\mytable{grid} demonstrates good performance for small grids where the
training data is large enough to accurately cover the state space. Note also
that the unregularized M-estimator performs the same as the original estimator
over all grid sizes. This is because the weighting matrix has no effect due to
the the equally likely transitions, which are already well-balanced. However, the role of regularization becomes more important as
the grid grows larger; this is because the fixed sample size leads observed
states to be more spread out and revisited less often.

\subsection{Chain configuration}

\begin{table}[!h]
\centering
\begin{tabular}{lllll}
\toprule
Reset probability & $\thetaspec$ & $\thetam$ & ($\thetam$,
$\lambda=1\textsc{e}\text{-}3$) \\
\midrule
0.1 & 0.80 & 0.73 & \textbf{0.72}\\
0.3 & 0.82 & 0.80 & \textbf{0.77}\\
0.5 & 1.24 & 0.96 & \textbf{0.69}\\
\bottomrule
\end{tabular}
\caption{Relative norm difference between estimated and true joint
probability, averaged over 100 test examples.}
\label{table:chain}
\end{table}

The chain configuration (\mysub{chain}) mimicks the chain problem in
reinforcement learning \citep{strens2000bayesian,poupart2006analytic}. Each
hidden state transitions to the next with probability $1-p$ and resets to the
first state with probability $p$. We use 50
training examples for each $p$.

As the reset probability increases, the data distribution becomes more heavy
tailed. This is reflected in \mytable{chain}, as the weighting makes a
larger impact over highly skewed distributions. As very few
examples are seen with the last few states, the $L_1$ penalty has a growing impact
as well.

\begin{table*}[!htb]
\centering
\begin{tabular}{llllllll}
\toprule
Data set & Type & Training set & \# Obs. states & $\thetaspec$ & $\thetam$ &
($\thetam$, $\lambda=1\textsc{e}\text{-}5$) &
$\hmbtheta^{\textsc{em}}$\\
\midrule
\emph{Alice} & Text & 50,000 & 26 & 0.22 & \textbf{0.20} & \textbf{0.20} & 0.14\\
\emph{Splice} & DNA & 100,000 & 4 & 0.41 & 0.40 & \textbf{0.35} & 0.19\\
\emph{Bach Chorales} & Music & 4,693 & 20 & 0.31 & 0.28 & \textbf{0.25} & 0.24\\
\emph{Ecoli} & Protein & 1,407 & 20 & 0.14 & \textbf{0.13} & 0.15 & 0.12\\
\emph{Dodgers} & Traffic & 30,000 & 10 & 0.42 & \textbf{0.38} & 0.39 & 0.33\\
\bottomrule
\end{tabular}
\caption{Predictive test error for three spectral estimators---\citet{hsu2012spectral}, M-estimator, and regularized M-estimator---and EM.  In many cases the M-estimators approach the performance of
  EM.}
\label{table:data}
\vspace{-1ex}
\end{table*}

\subsection{Synthetic \acrlongpl{HMM}}
\label{subsec:synthetic}
We generate two large synthetic data sets following well-behaving
\glspl{HMM}: one system uses $m=5$ hidden states and $n=10$ observed
states, and the other uses $m=10$ hidden states and $n=20$ observed
states. We perform both full rank and low rank estimation over $10,000$
training examples and analyze held-out prediction error.

\begin{figure}[tb]
\centering
\includegraphics[width=0.7\columnwidth]{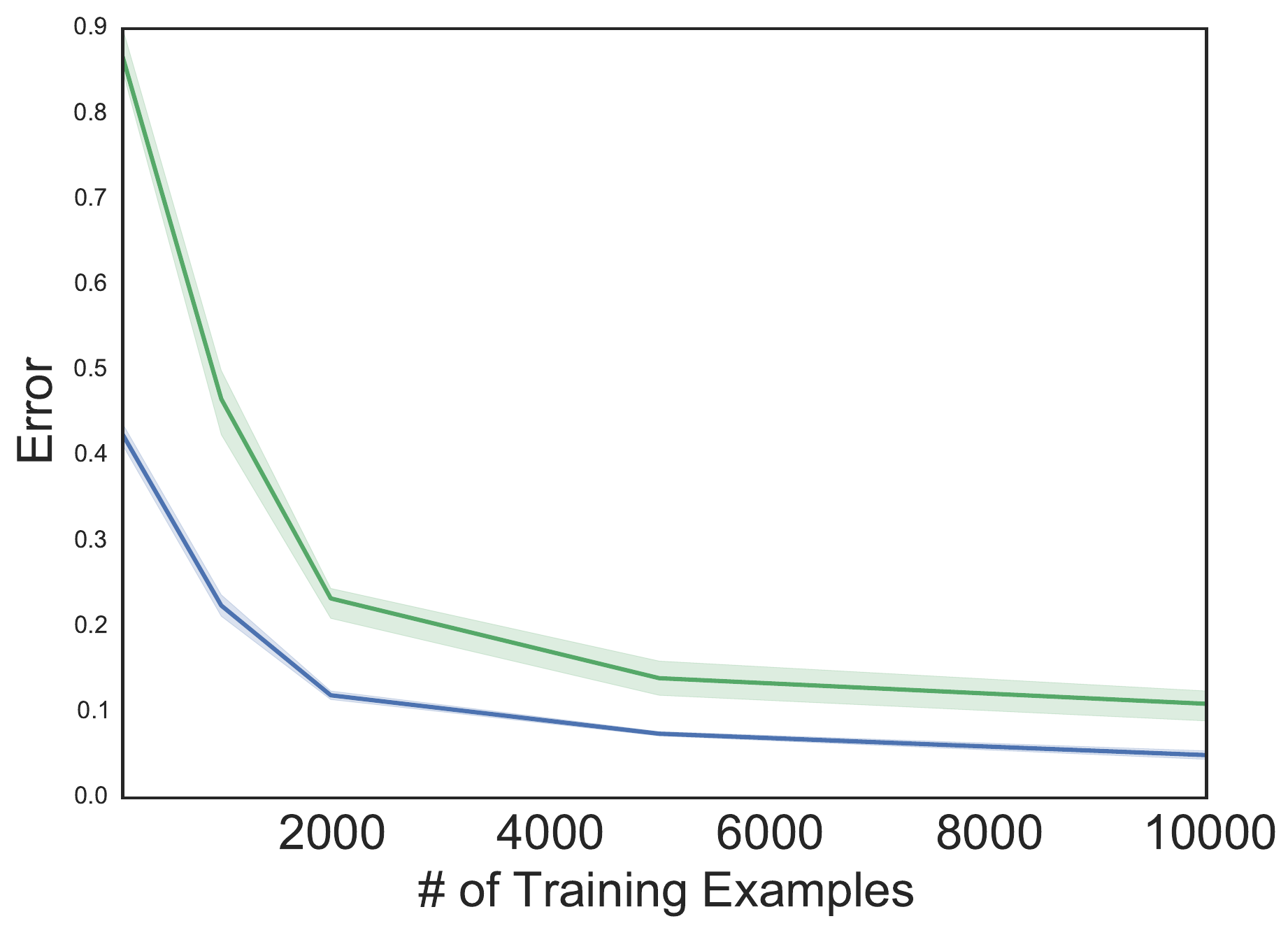}
\includegraphics[width=0.7\columnwidth]{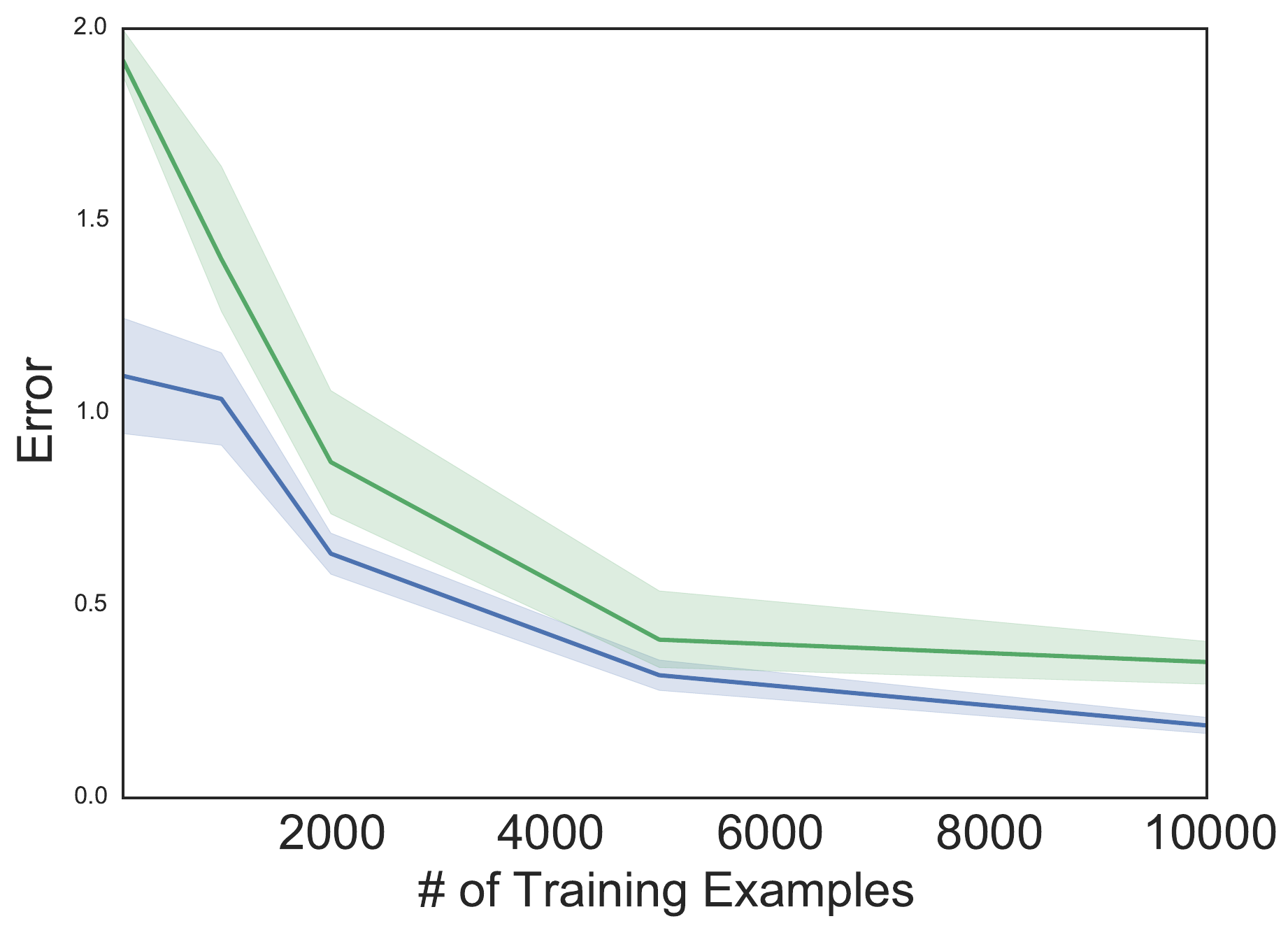}
\caption{%
Predictive accuracy of original
estimator (\green{green}) and M-estimator (\blue{blue}) over \# of training examples, with standard error bars
taken over 100 simulations. Top: $m=5$ hidden states with $n=10$
hidden states.
Bottom: $m=10$ hidden states with $n=20$ observed states.
}
\label{fig:synthetic}
\vspace{-2ex}
\end{figure}

In \myfig{synthetic}, we see that with few training examples, the
M-estimator's optimal weighting scheme is crucial for reasonable
performance.  Moreover, as explained in theory, the variance of the
M-estimator is much lower than the original spectral
estimator. The original estimator and the M-estimator converge at the
same rate and eventually reach competitive errors. However, the
M-estimator achieves this much faster in practice even in these
well-behaving dynamical systems.

\subsection{Real data sets}

We now examine the performance of the estimators for 5 separate data
sets: in the \emph{Alice} novel available in Project Gutenberg, the
task is to predict characters after having trained over the first
50,000 of them; the \emph{Splice} data set consists of 3,190 examples
of DNA sequences which have length 60 and the task is to predict the
remaining A,C,T, or G fields; the \emph{Bach Chorales} consists of
discrete event sequences in which the task is to predict the correct
pitch of melody lines; \emph{Ecoli} describes sequencing information
of protein localization sites; \emph{Dodgers} examines link counts
over a freeway in Los Angeles. These last four data sets are available
from \citet{lichman2013uci}.

\mytable{data} indicates the average prediction error on held out
data.  The results are consistent with that of the toy configurations
and synthetic benchmarks. In all data sets, the M-estimator surpasses
the original estimator. The benefit of sparse regularization tended to
vary, as we did not choose to tune this hyperparameter per data set.
We also compared to EM with random initializations as a benchmark to
likelihood-based methods. Many local optima performed poorly; the best
solutions found after enough random initializations uniformly
performed better than the spectral estimators over all data sets.

\section{Discussion and Related Work}
\label{sec:discussion}
In this work, we focused on the application of M-estimation to
estimating parameters of \glspl{HMM}. Our analysis
and algorithms carry over almost identically for predictive
state representations (e.g. in
\citet{siddiqi2010reduced,song2010hilbert,boots2010closing}).
Estimating parameters for other latent variable models can also be
easily formulated as generalized method of moments problems. For
example,
following \citet{anandkumar2012method},
a mixture model specified by $\Pr(h = j) = \omega_j$ and
$\Pr(x = i\mid h =j)=\mbM_{ij}$ for $i\in [n], j\in[k]$,
has moment conditions
\begin{align*}
m_{1}(\mbM,\omega)
&
=
\bP_{2,1} - \mbM\operatorname{diag}(\omega) \mbM^\top
,
\\
m_{x}(\mbM,\omega)
&
=
\bP_{3,x,1} - \mbM\operatorname{diag}(\mbM^\top
e_x)\operatorname{diag}(\omega) \mbM^\top,
\end{align*}
for all $x\in[n]$, where $e_x$ is the unit vector equal to one at
index $x$.
Closest to our approach is that of \citet{kulesza2015lowrank}, who
propose a weighting scheme to address fundamental issues with
low rank spectral learning. Their weighting scheme can be seen as
redefining the moment conditions
\begin{equation*}
\vspace{-0.25ex}
m_x(\mbtheta) = \bP_{3,x,1} - \B_x\W\bP_{2,1}
\quad \forall x\in[n].
\vspace{-0.25ex}
\end{equation*}
With this moment condition, solvers using singular value decomposition
avoid instabilities as noted in \citet{kulesza2014lowrank}.  In contrast, our
\gls{GMM} approach takes the direct path of
weighting the moment conditions, i.e., the error in the statistics for
estimating the moments. \citet{kulesza2015lowrank} also
require that a domain expert specify the weighting matrix $\W$; our
$\thetam$ is automatically given by our optimal choice of weighting
matrix.  That said, in situations where domain experts can connect a
choice of $\W$ to a specific task, one can forgo
sample efficiency and specify the weighting matrix of the \gls{GMM}
manually.

Also related to our work are methods that use spectral methods to
initialize techniques for maximum likelihood estimation
\citep{zhang2014spectral,balle2014methods}. \citet{shaban2015learning}
follow this approach and propose a two-stage procedure, which
corresponds to typical spectral estimation in the first stage and
optimization upon the second to ensure feasible solutions (which our
method does not).  While we also have an iterative procedure that
begins with a spectral initialization, each of our steps is still
within the spectral framework.  Our approach of weighting the moments
and considering suitable penalization is orthogonal to the use of the
spectral estimates for initializing other estimation techniques. It
remains open to explore the benefits of these approaches when merged
in practice.

To our knowledge, our work is the first to achieve optimal sample
efficiency rates for spectral estimation, and we provide a principled
approach to incorporating regularization into the process.  However,
we now have a highly nonconvex optimization problem, and we also rely
on row-level elements of the data.  Addressing these concerns, while
maintaining sample-efficiency and accuracy bounds, remains an
important direction for future work.

\subsubsection*{Acknowledgements}
We thank Maja Rudolph and Dawen Liang for their helpful comments and
support from NSF ACI-1544628.

\section*{References}
\renewcommand{\bibsection}{}
\bibliographystyle{apalike}
\bibliography{aistats2016}

\clearpage
\appendix

\section{Proof of Proposition 1}
\begingroup
\def\thetheorem{1}
\begin{proposition}
\propequivalence
\end{proposition}
\addtocounter{theorem}{-1}
\endgroup

\begin{proof}
Let $x\in[n]$, and consider a solution to
the moment conditions
for parameter
$\B_x\in\mathbb{R}^{n\times n}$ given by
\begin{equation}
\label{eq:hmmrank}
\operatornamewithlimits{min}_{\B_x}
\|\bP_{3,x,1}- \B_x\bP_{2,1}\|_F^2
\end{equation}

\myeqp{hmmrank} can be solved using any convex program, or, by
the Eckart-Young theorem \citep{eckart1936approximation}, through singular
value decomposition. Thus we recover the original spectral estimator: \myeqp{hmmrank} is equivalent to a
singular value decomposition as standard methods in spectral learning do
\citep{hsu2012spectral, boots2010closing, boots2011online, huang2013fast}. Note
further that while this problem is nonconvex, all local optima are also global
\citep{nati2003weighted}. Hence the estimates we obtain using optimization
routines are consistent.

\citet{hsu2012spectral} derive
\myeqp{hmmrank} from a different standpoint and consider the special case of
full rank $k=m$.  They proceed to relax the rank constraint by observing that the
parameters are learned up to a similarity transform: given the triplet
$(\b_1, \{\B_x\}, \b_\infty)$ and an invertible
matrix $\S\in\mathbb{R}^{n\times n}$, the transformed triplet $(\b'_1= \S
\b_1, \{\B'_x=\S\B_x\S^{-1}\},
\b'_\infty= \S^{-T} \b_\infty)$ provide the
same joint probabilities as written in
Equation (5).

Instead of choosing an invertible similarity transform, one can
find $\U\in\mathbb{R}^{n\times k}$ such that $\U^\top\bP_{2,1}$ (equivalently,
$\U^\top\O$) is invertible, as any inversions regarding $U$ are only involved through the
product $\U^\top\bP_{2,1}$.
A natural choice is to let $\U$ be the matrix of $k$ left-singular vectors of
$\bP_{2,1}$ \citep[Lemma~2]{hsu2012spectral}. Then an equivalent optimization
procedure to Equation
\ref{eq:hmmrank} is simply
\begin{equation}
\label{eq:frobenius}
\min_{\B'_x} \|\bP_{3,x,1} - \B'_x\bP_{2,1}\|_F^2
\end{equation}
where
$\B'_x\equiv \U^\top \B_x (\U^{T})^\dagger=(\U^\top\O)\A_x (\U^\top\O)^{-1}\in\mathbb{R}^{k\times k}$.
The advantage is that $\B'_x$ is automatically constrained to be of rank $k$
through the similarity transform on $\A_x$ given by $\U^\top\O$. This can be solved trivially with
$\B'_x=\bP_{3,x,1}\bP_{2,1}^\dagger$, and in terms of the original parameter
$\B_x=(\U^\top \bP_{3,x,1})(\U^\top\bP_{2,1})^{-1}$ \citep[Proof of Lemma 3]{hsu2012spectral}.
\end{proof}

\section{Proof of Proposition 3}
\begingroup
\def\thetheorem{3}
\begin{proposition}
\propgradient
\end{proposition}
\addtocounter{theorem}{-1}
\endgroup

\begin{proof}
For a general quadratic matrix function $f(\mbtheta) = y(\mbtheta)^\top \W y(\mbtheta)$ with given matrix $\W $, its gradient is
\begin{equation*}
\nabla f(\mbtheta) = [\nabla y(\mbtheta)]^\top (\W  + \W ^\top) y(\mbtheta)
\end{equation*}
Hence for our situation where $\W $ is symmetric, it is
\begin{align*}
\nabla_{R} \mathcal{L}
&=
2\left[\nabla_{R}\left[
[\widehat P_{3,x,1}]_{ij} - [R_x]_{i\cdot}S_x^\top[P_{2,1}]_{\cdot j}
\right]_{xij\in[n^3]}\right]^\top
\\
&\quad
\W
\left[
\widehat m_{xij}(\mbtheta)
\right]_{xij\in[n^3]}
\\
&=
2\mathcal{J}_R^\top \W
\left[
\widehat m_{xij}(\mbtheta)
\right]_{xij\in[n^3]}
\end{align*}
The Jacobian $\mathcal{J}_R$ is a $n^3\times n^2k$ matrix, with elements $(xij,
uvw)\in[n^3]\times [n^2k]$. The $(xij,uvw)^{th}$ entry is the partial derivative
of the $xij^{th}$ moment $\widehat m_{xij}$ on $[R_u]_{vw}$:
\begin{align*}
[\mathcal{J}_R]_{xij, uvw}
&=
\frac{\partial}{\partial [R_u]_{vw}}
\left[
- \sum_{r=1}^k [R_x]_{ir}[S_x^\top]_{r\cdot}[P_{2,1}]_{\cdot j}
\right]
\\
&=
\begin{cases}
-[S_x^\top]_{w\cdot}[P_{2,1}]_{\cdot j} &\text{ if }x=u,~i=v\\
0 &\text{ otherwise}
\end{cases}
\end{align*}
Similarly, there is a Jacobian $\mathcal{J}_S$ when taking the gradient with respect to $S$, and
by the same logic the Jacobian with respect to $S$ is
\begin{align*}
[\mathcal{J}_S]_{xij, uvw}
&=
\frac{\partial}{\partial [S_u]_{vw}}
\left[
- \sum_{s=1}^n \sum_{r=1}^k [R_x]_{ir}[S_x]_{sr}[P_{2,1}]_{sj}
\right]
\\
&=
\begin{cases}
-[R_x]_{iw}[P_{2,1}]_{vj} &\text{ if }x=u\\
0 &\text{ otherwise}
\end{cases}
\qedhere
\end{align*}
\end{proof}

\end{document}